# Low-pass Genomewide Sequencing and Variant Imputation Using Identity-by-descent in an Isolated Human Population

Gusev A[1,2], Shah MJ[1,3], Kenny EE[1,4], Ramachandran A[1,2], Lowe JK[4,5,6], Salit J[4,7], Lee CC[3], Levandowsky EC[3], Weaver TN[3], Doan QC[3], Peckham HE[3], McLaughlin SF[3], Lyons MR[3], Sheth VN[3], Stoffel M[8], De La Vega FM[3], Friedman JM[4], Breslow JL[4], Pe'er I[1,2]

[1]Department of Computer Science and [2]Center of Computational Biology and Bioinformatics, Columbia University, New York NY; [3]Life Technologies, Inc., Beverly, MA; [4]The Rockefeller University, New York, NY; [5]Department of Molecular Biology, Massachusetts General Hospital, Boston, MA; [6]Program in Medical and Population Genetics, The Broad Institute of Harvard and MIT, Cambridge, MA; [7]School of Medicine, Cornell University, New York, NY; [8]Institute of Molecular Systems Biology, Swiss Federal Institute of Technology (ETH), Zurich, Switzerland;

## Abstract

*Whole-genome sequencing in an isolated population with few founders directly ascertains variants from the population bottleneck that may be rare elsewhere. In such populations, shared haplotypes allow imputation of variants in unsequenced samples without resorting to statistical methods, as in studies of outbred cohorts. We focus on an isolated population cohort from the Pacific Island of Kosrae, Micronesia, where we previously collected SNP array and rich phenotype data for the majority of the population. We report identification of long regions with haplotypes co-inherited between pairs of individuals and methodology to leverage such shared genetic content for imputation. Our estimates show that sequencing as few as 40 personal genomes allows for imputation in up to 60% of the 3,000-person cohort at the average locus. We ascertained a pilot data-set of whole-genome sequences from seven Kosraean individuals, with average 5X coverage. This dataset identified 5,735,306 unique sites of which 1,212,831 were previously unknown. Additionally, these Kosraen variants are unusually enriched for alleles that are rare in other populations when compared to geographic neighbors. We were able to use the presence of shared haplotypes between the seven individuals to estimate imputation accuracy of known and novel variants and achieved levels of 99.6% and 97.3%, respectively. This study presents the first whole-genome analysis of a homogenous isolate population with emphasis on rare variant inference.*

# Introduction

Founder populations play significant roles in population genetics and trait mapping due to the effects of bottlenecks and drift on their genetic variation[1]. Such populations are singularly useful in identifying rare disease variants that often appear in the isolated cohort at a higher frequency or within a more clearly discernable haplotype structure[2] than in out-bred populations. Additionally, identified variants are still valuable beyond the isolated group as their effect replicates in more outbred populations[3,4] and can implicate new functionally important genes. Next generation sequencing of personal genomes has revealed a multitude of rare variants. Presently, high-quality personal genomes exist from representatives of the major continental groups[5,6,7,8,9,10,11], but only genotyping and lower-throughput sequencing data is available for isolated populations[12,13]. This paper reports low-pass whole-genome sequencing and analysis of seven individuals from an isolated Pacific population, chosen specifically for the insight they might provide into the larger cohort as well as the presence and functional importance of rare variants they carry.

The cost of whole-genome sequencing is not trivial and the best strategy for identification of rare causative variants must balance the number of genomes sequenced with the insights gained that are applicable to different populations and multiple traits. For common traits, one may sequence a reference panel to statistically impute variants in populations represented by such a panel[14]. However, this requires sequencing high numbers of genomes and is still severely underpowered in populations or variants that are underrepresented in such datasets (e.g. isolated populations[15] and rare variants[16]). For Mendelian diseases a successful strategy has been whole-exome capture in a small number of individuals[17,18]. However, such studies are limited to extremely penetrant phenotypes, inherently ignore non-coding regions, and do not yet scale to population-based analysis[19]. Another alternative strategy has been targeted re-sequencing of candidate loci detected in a GWAS across many individuals. Nevertheless, pursuing such a strategy genome-wide is still resource intensive despite a considerable drop in sequencing costs, and scales poorly for multiple traits across a large number of loci in each.

We set out to leverage the opportunities and address the challenges of sequencing-based mapping in a multi-trait GWAS cohort from an isolated population. Ongoing work by large sequencing consortia, such as the 1,000 Genomes Project[20,21], has shown that analyzing multiple individuals, even at low coverage, improves quality and completeness of detecting and calling novel variants. Moreover, information from a small sub-sample of sequenced individuals combined with relatively inexpensively acquired SNP array platforms can be used to impute much of the missing variation with high accuracy[14,22]. In the current study we used this knowledge to develop a sequencing-based framework that leverages the inherent potential of a sizeable phenotyped cohort with a small founder population. We applied it to the Kosraen data set in which we previously found an abundance of long stretches of the genome identical by descent even between reportedly unrelated pairs of individuals[23].

We shot-gun sequenced a pilot group of seven individuals and performed multi-sample calling and imputation to quantify the informativeness of this cohort. The detected variants were validated and compared with those observed in other published whole-genome sequencing efforts from different populations. Internally, we analyzed the distribution of all variation as well as individual functional classes. Lastly, we estimated the effectiveness of IBD segments detected from SNPs in predicting the underlying untyped variants.

## Results

We have been studying genetic determinants for a multitude of traits in a cohort of 2,906 individuals (the majority of adults) from the Micronesian island of Kosrae. This cohort has been previously genotyped on the Affymetrix 500k SNP array platform, reporting positive GWAS results for seven phenotypes (including details on the screening and genotyping process[24], and statistical analysis of traits within the population[4]). Subsequently, we reported a GWAS in which 27 traits were analyzed under family-based models[25] and quantified the abundance of IBD segments within the cohort[23].

We utilized the autosomal SNP genotype data to estimate pervasiveness of IBD in genomic regions between arbitrary pairs of samples, and as a consequence, the potential for IBD-based imputation in this population. Regions of IBD were recovered using GERMLINE, a tool for efficient whole-genome IBD detection from partially phased data[23] (Materials & Methods). For the purpose of imputation, we conservatively examined only IBD segments longer than 5cM, where GERMLINE has been demonstrated to have 100% specificity in simulation[23,26]. We found that for an average individual, such regions span a total of 10.8% of all genotypes in the remaining cohort. We then seek to estimate the utility of these IBD segments for imputing genomic data within the population from a sequenced subgroup. We first make a simplifying assumption to trust all of the detected IBD regions and ignore subsequent potential sequence error or variant differences. In other words, we presume that a sequenced individual can infer all information in IBD regions it shares with other un-sequenced samples, which yields an upper-bound estimate of imputation capacity. We developed a novel method for optimizing the selection of highly representative individuals to sequence and quantifying the amount of data that can be inferred from their genomes. Briefly, the method uses a special-purpose data structure, an interval tree, to find the individual with highest totality of IBD sharing on both homologous copies of the genome; isolate that individual into the sequence panel; and, after excluding all of their shared segments, continue the calculation in an iterative manner (Figure 1A). This method, INFOSTIP (http://www.cs.columbia.edu/~itsik/Software.htm) is discussed in greater detail and evaluated in Supplementary Methods. Figure 1B presents the results of this analysis, an estimate of the fraction of the genotyped sample cohort that can be inferred as a function of the sequencing budget (number of sequenced individuals). We observe that sequencing 50 randomly chosen individuals (1.7% of cohort) would give us the potential to impute both alleles of variants in 59.5% of the cohort genome, but choosing individuals in an optimized fashion using INFOSTIP decreases the sequenced sample size needed for the same benchmark by 24% to 38 individuals (1.3% of cohort). Remarkably, sequencing only seven individuals (0.24% of cohort) still provides imputation capacity of 24% of the cohort genome. For comparison, we conducted the same analysis within a cohort of 1,200 Ashkenazi individuals[27], a population known to be isolated but less densely related. In this case we found that utilizing our optimal selection method and sequencing 38 individuals gave us the potential to impute variants in only 16% of the cohort genome, whereas sequencing seven individuals allowed us to infer only 4% of the cohort genome (see Figure 1B, Supplementary Figure 1 and Materials & Methods for additional analysis). This type of imputation is agnostic of allele frequency as long as a relevant IBD segment is available.

*Sequencing benchmarks*

We sequenced a discovery panel of seven low-pass personal genomes, four of which were selected according to the aforementioned procedure with the remainder chosen according to phenotype (Materials & Methods). For each of the seven individuals, 10-30 micrograms of genomic DNA was used to generate a library following Life Technologies' long mate-pair protocol. The libraries were sequenced using the SOLiD$^{TM}$ System, with 8,239,389,322 raw 50 bp mate-paired reads and an additional 740,209,937 raw 35 bp mate-paired reads, generating a total of 438 Gb. The raw reads were aligned and paired to the reference human genome (hg18) using the AB SOLiD Corona Lite pipeline (http://solidsoftwaretools.com). Up to 3 mismatches were allowed for 35 bp reads and up to 5 mismatches were allowed for 50 bp reads. This generated 158 Gb that map to the genome as uniquely placed normal mate pairs within the expected distance (1.5 kb insert size), order and orientation. 96.6 Gb of these uniquely placed normal mates are non-redundant, which represents a >30X coverage of the "Kosraen genome". On average, 3-6X sequence coverage of non-redundant normal uniquely placed pairs was achieved for each individual (Supplementary Table 1).

*Variant calling*

Following the structure of the 1,000 Genomes Project low-pass pilot, we performed variant calling on all seven samples together using the Genome Analysis Toolkit[28] as well as several steps of imputation. In summary, we performed local realignment and quality score recalibration of the reads from each individual separately; variants in all samples were then called together using an iterative Bayesian algorithm that attempts to infer allele frequency in the population in support of individual genotype calls; for previously known variants, we used a strict call quality threshold to minimize false-positives; for novel variants, we performed an additional variant quality score recalibration procedure to minimize expected false-positives; lastly, we performed internal imputation using the BEAGLE framework[29] and external imputation to the 1,000 Genomes pilot haplotypes using MaCH[22]. This strategy allows us to leverage the presence of a confidently observed variant in higher-coverage samples to recover calls in lower-coverage samples that would not have been called individually. The detailed variant calling protocol is described in full in the Materials & Methods.

We performed rigorous quality control on the set of called variants using the available array-based genotypes and additional novel genotyping as validation data (Table 1). Of the previously known sequence calls, 2,958,772 overlapped with the genotyped variants and were used for validation, allowing evaluation of specificity of detecting non-reference sites, as well as of calling each genotype class. We measured the specificity of calling SNPs from all seven samples together on non-reference calls at these known sites to be 98.2%. We independently computed similar levels of specificity, 98.9%, that would be expected based on the observed transition/transversion ratio of 2.07 across all of the previously known variant sites (see Materials & Methods). Per-call specificity of SNP calling from the samples together varied by sample, with an average of 94.1% for heterozygous calls and 92.5% for homozygous calls (Supplementary Table 2). Low coverage often caused heterozygous SNPs to be identified as homozygous. The low coverage from this pilot, our strict quality thresholds and the pooling of all seven samples led to a sensitivity to detect non-reference sites of 92.8%. For variants that were not previously known, we observed an overall transition/transversion ratio of 1.74, corresponding to expected specificity of 88.9%. We validated a total of 64 called novel sites using Sequenom genotyping (Materials & Methods, Supplementary Table 3) and found the empirical specificity of non-reference calls to be 87.5%, in-line with our overall estimates. Additionally, we detailed the array-based validation results at each step of the calling pipeline and found the largest increase in accuracy to come from calling all samples together rather than individually and from internal imputation (Supplementary Table 2) as previously reported [21,30,31].

*Variation discovered*

We now focus specifically on variants identified in the autosomes, as these are directly applicable to our IBD-based analysis. Our final set of SNVs contained 22,221,159 non-reference calls across all seven discovery samples for a total 5,735,305 unique sites of which 1,212,831 (21%) were previously unknown (not in dbSNP

v130). The total number of non-reference calls ranges across individual samples from 3.1 to 3.4 million (Supplementary Table 4). We expect this to be an incomplete estimate, representing the limitation of low-pass sequencing in calling variants at low-coverage sites due to undersampling of the variant allele. For a fair comparison to other genomes, we extrapolate the total number of variants in the mappable genome based on the error rates described in the previous section (see Materials & Methods). Thus, we estimate an average Kosraen sample to contain 3,241,030 total autosomal variants (±66,996 s.d.). Comparing the genomewide estimates to a variety of published genome sequences (Supplementary Figure 2), we find the overall number of variants is nearly identical to the 3.25 million observed in average East Asian autosomes[5,7,10]. Comparing the called variants to markers previously annotated in dbSNP v130, we estimate 10.2% (0.90% s.d.) of the variants in the average Kosraen to be novel, not significantly different from the 9.4% average (3.5% s.d.) in other East Asian samples[5,7,10]. Due to the long history of isolation in this cohort we suspect many of the observed novel variants to be mutations private to the island.

Within an average Kosraen sequence, we find 50.7% of non-reference sites to be homozygous (±2.2% s.d.). We caution that this figure is likely to reflect under-called heterozygotes and expected to drop with deeper sequencing[11]. However, it is significantly higher than the observed homozygosity rate in the other personal genomes, even those with similar coverage (next highest - 41.7% in Anonymous Asian[10]). Historically, the population experienced a series of severe bottlenecks, which would have resulted in many variants drifting to higher frequency and becoming homozygous.

We estimate the unique novel variation that each sequenced individual contributed by averaging over all 5,040 permutations of the seven samples (Supplementary Figure 3). We find that the first Kosraen sample to be considered contributes approximately a third of the novel sites that were called in all samples (352,162 of 1,014,310; Supplementary Table 5), with decreasing contribution of subsequent samples. This decrease, initially by 40% of novel variants that are overlapping between an average pair of samples, becomes more gentle for additional samples, reflecting the enrichment of rare variants in this set, eventually reaching approximately 50,000 variants contributed by the last sample only - the average number of single-carrier variants.

*Analysis across sequenced populations*

We analyzed the population specificity of the Kosrean variants by examining their respective allele frequency in the reference populations sequenced as part of 1,000 Genomes Pilot I. Within each of the pilot cohorts of Yoruban (YRI), European (CEU), and East Asian (JPTCHB) origin, we measured the allele frequency of homozygous variants from the Kosraen (KOS), Korean (SJK), European (JCV), and Yoruban (YRI) sequenced genomes. The proportion of variants that fall into each allele frequency window of a reference cohort is shown in Figure 2. Focusing on the differences between the Kosrean and their closest analyzed neighbor - the SJK Korean genome - we observe that the average Kosrean is relatively enriched for rare variants in all three populations. Specifically, the percentage of alleles in an average Kosraen that were uncommon in JPTCHB (below 10% frequency) was 4.4-fold higher than that of SJK (Figure 2A). In the other cohorts, we see more subtle but consistent enrichment of 1.42-fold and 1.24-fold in uncommon CEU and YRI alleles respectively (Figure 2B,C). This trend suggests that lower frequency alleles in other populations that are present in Kosrae have drifted to higher frequency within the cohort as compared to the Korean genome.

*Analysis within the isolate population*

We annotated the called variants according to their functionality and analyzed the carrier frequencies of sites that have coding or splicing implications. From the overall frequency distributions (Supplementary Figure 4), we observe a significant increase in novel coding variants appearing as either singletons or fixed non-reference alleles when compared to novel non-coding variants ($P=8.7 \times 10^{-10}$ from $\chi^2$ test). Focusing on specific coding sub-classes in Figure 3, we examine variants called in all samples and annotated as: (**A**) splice junction, all coding, synonymous, missense and nonsense; as well as (**B**) showing all known and novel variants (full data in Supplementary Table 5, Supplementary Figure 4). As reported in previous studies [17,32,33], we see significant over-abundance of singleton nonsense mutations compared to other singleton variant classes ($P=5.4 \times 10^{-4}$ or $1.4 \times 10^{-4}$ compared, respectively, to coding or all variants). This over-abundance is consistent with the effects of purifying selection negatively affecting the frequency of functionally important variants. All of the detected non-synonymous mutations were significantly enriched for genes with the gene ontology term 'olfactory receptor activity' ($P=4.4 \times 10^{-7}$ after Bonferroni correction[34]; increased enrichment when compared to synonymous mutations), evidence of a continued process of pseudogenization in this family.

*Structural variation*

We identified short insertions and deletions using the Dindel algorithm[35] in all samples together. Briefly, Dindel identifies candidate indels within the read data and then attempts to align them to haplotypes that represent alternative sequences to the reference (detailed protocol in Materials & Methods). Supplementary Figure 5 details the distribution of novel and previously known indels across the seven sequenced individuals. Overall, we observe a steep decrease of indel carrier rate, with 46% of all indels present in a single individual. As with SNVs, the novel indels tend to be enriched for singleton and fixed indels in this cohort when compared to previously known sites.

We identified structural variants longer than 10kb using the SOLiD Software Tools, which combines depth-coverage, predicted mappability, and GC-content, within a Hidden Markov Model framework to make CNV region calls. Overall, an average individual contained 77.1MB of copy-variable regions end-to-end, with the longest variant being a 7.6MB heterozygous deletion on chromosome 19p13. We analyzed the lengths of the CNVs found in all the samples by variant type and length (Supplementary Figure 6). In particular, CNVs of size less than 100 kb constitute 66.9% of the calls, with most being heterozygous deletions. We also looked at the number of shared and private CNVs among the Kosraen individuals, with a CNV being considered shared between two individuals if the overlap between the two called regions was at least 80%. For an average individual, approximately 20% of CNVs are shared by the entire population.

*IBD analysis*

Assessing the IBD-based motivation for this pilot, we focused on the 1,522 shared segments predicted between the sequenced individuals, ranging in length from 330kb to 74Mb. Unlike the conservative INFOSTIP analysis, which examined fewer but higher-quality IBD segments, these segments were detected using GERMLINE's default parameters with no adjustment (3cM segment length minimum), allowing us to estimate IBD accuracy under practical conditions. We evaluated the accuracy and utility of the IBD-based approach by examining variant concordance within these regions. Specifically, two samples that are IBD across a region should not have sites with homozygous calls for opposite alleles in that region. For a pair of such samples, we examine all sites in the IBD region that are mutually homozygous with at least one sample being non-reference, and report concordance as the percentage of these sites that are not homozygous for opposite alleles. Lack of such concordance is indicative of either falsely called IBD or poor genotype calls due to under-sampling of sequence reads (true heterozygous sites miscalled as homozygous). Aside from some effects on multi-sample calling, the concordance rate can be treated as a measure of baseline homozygous variant imputation accuracy when one of the individuals had not been sequenced. Figure 4 shows concordance across IBD segments, separated into previously known (**A**) and novel (**B**) variants. For comparison, we measured the background distribution such concordance across 30 random selections of same-sized regions, shown in black points. We observe the vast majority of IBD segments having nearly 100% concordance, with only 1.4% and 10.6% of segments below 90% concordance for known and novel variants, respectively. If we take a weighted average across all segments, the aggregate concordance is 99.6% (known) and 97.3% (novel) in IBD-segments, providing encouraging estimates for accuracy of IBD-based imputation. This is compared to a background concordance statistic averaging 82.9% (known) and 31.0% (novel) in permuted segments. We attribute the difference between known and novel concordance in IBD regions to be an artifact of lower sensitivity to novel variants and the overall deviation from full concordance to be indicative of inaccurate detection of IBD regions or their exact boundaries.

Of particular interest to the IBD community is the minimum length at which stretches of SNPs identical by state (IBS) are still predictive of identity at un-typed variants[36]. To estimate this, we omit the minimum segment length restriction for IBD detection, resulting in a tally of all runs of at least 128 IBS SNP-array sites in the sequenced samples, rather than the set of putative IBD regions we considered thus far. We measured concordance in length windows of 1cM from 0-10cM and above. Supplementary Figure 7 shows this concordance distribution for known and novel variants, as well as the number of segments measured within each window. As previously documented[23], we see a direct correlation of concordance with segment length, as longer IBS segments are more likely to represent true recent IBD. However, we observe only a slow decrease in concordance from high-quality 10cM segments down to 2-3cM indicating either a small number of false-positive segments or overcalled false IBD primarily around the boundaries of true IBD segments. Even within the 0-1cM length window (median physical length 815kb) we see 98.9% (known) and 91.6% (novel) concordance, significantly above the average in non-IBD regions. These initial findings suggest that even very short IBS segments can be useful for variant inference.

## Discussion

While the population genetics of isolated groups has been of interest for decades, the contribution of such groups to understanding heritable traits is strongly dependent on the research methodology employed. In the context of inbred populations, linkage analysis of Mendelian traits using microsatellite scans has mapped many mutations that are rare in the general population. In contrast, association analysis with SNP arrays relies on linkage disequilibrium in populations, and by primarily targeting common, ancient variation has been mainly applied to outbred peoples. High-throughput sequencing now makes possible discovery of rare variants in the general population but, as shown in this paper, with the proper strategy can be applied to the study of isolated communities efficiently and to great effect.

Different strategies for high throughput sequencing offer various tradeoffs of investment and potential for discovery. Whole-genome sequencing at high coverage is the gold standard, but is still expensive to pursue with substantial sample sizes. Focusing on a captured target, either around a genomic area of interest[37] or considering all exonic regions[38] sacrifices potential information from most of the genome for high quality data regarding the most promising parts. Low pass sequencing offers a different tradeoff, considering the entire genome, but accepting lower-quality data. Indeed, the first reference and multiple personal genomes[8,11,39] are all low-pass, with meaningful insights regarding technology[40], population genetics[41], and mutation detection[17,42]. This work follows suit and provides population-based sequencing of Pacific Islanders.

With an emphasis on accurate variant detection, we ascertained the effectiveness of low-pass sequencing in conjunction with multi-sample calling, achieving overall non-reference specificity and sensitivity above 90%. In particular, some of the lower-coverage samples netted 2-3 fold increases in accuracy when compared to independent calling. Overall, this strategy allowed us to uncover 1,212,831 previously unknown variants with high specificity.

Examining the spectrum of variation, we explored characteristics unique to this cohort, which had undergone a series of severe bottleneck events. As expected from such an extreme founder population, the qualitative variant statistics reveal an abundance of novel variation and overall homzoygosity. Moreover, those sites that have been observed in other sequenced populations still exhibit enrichment for alleles that are rare outside of Kosrae. Demonstrating the effects of purifying selection, we observe a significant abundance of rare coding variants and singleton nonsense mutations compared to all variants and synonymous mutations, respectively.

Leveraging the wealth of relatedness and haplotype sharing in the population, we find 97.3% concordance of novel variants within segments shared IBD by the sequenced samples, demonstrating the potential for inferring such variants in other un-typed but IBD individuals. With a high rate of concordance even in very short putative IBD segments, we expect a full panel of 40 sequenced individuals to infer at least 60% of the overall population genome.

Our work highlights the manageability of population sequencing for isolated populations. While infrastructure efforts by large consortia such as the 1000 genomes lay foundations for comprehensive catalogs of variants in outbred populations, we demonstrate sequencing at the scale of an individual lab as a means to make genetics of such populations fully tractable.

As sequencing studies expand geographically to capture the bulk of common variation, isolated populations can help broaden our understanding of rare alleles. While this effort sequenced only a handful of individuals and the sequence coverage of each of them is low, their relation to one another and with many other islanders facilitates both reliable variant calling as well as powered association analysis to variants detected by full sequencing. This approach avoids the ascertainment bias of previous SNP-based studies, and suggests a strategy to leverage SNP array data in large samples, where sequencing is still expensive.

# Materials & Methods

*Sample selection*

Four of the samples (K1955, K2033, K5866, K1674) were selected using the INFOSTIP strategy to maximize their IBD-based inference capacity to the rest of the 2,906 genotyped individuals from the Kosraen cohort. The remaining three samples (K6169, K6494, K5675) were selected based on being phenotypic extremes for several metabolic traits, and as carriers of haplotypes associated with these traits (not shown).

*IBD segment analysis and imputation*

The pedigree of 2,906 Kosraen individuals was divided into three groups without replacement: two parents and a single child (trio), a single parent and a single child (duo), and single samples (unrelated). Using the BEAGLE framework[29], the individuals were phased and missing data inferred taking into consideration their respective group structure. The phased genotype data was processed with GERMLINE under default parameters and with genetic distance annotation data corresponding to the Affymetrix 500k chip to generate the genotype-based IBD shared segments. The same data was additionally processed with GERMLINE under the phase-specific haplotype-extension parameters, which explicitly treats each homolog separately in generating matches.

The INFOSTIP analysis was performed on both genotype and haplotype oriented IBD segments. For haplotype data, INFOSTIP executed upon each homolog as if it were an independent set of shared segments, but in choosing a sample for the sequence panel excluded all of the matches originating from that individual on either homolog. As such, a site must be either autozygous, or contained within an IBD segment of two differing sequenced individuals to be fully inferred. For genotype data, INFOSTIP ran with no modification and hence, a site is considered fully inferred if either homolog is in IBD with a sequenced individual. Because the imputed regions were SNP-chip oriented, the total cohort genome length was calculated as the individual end-to-end length of the genome that contained SNPs, multiplied by the number of samples (for genotype data) or twice the number of samples (for haplotype data). Supplementary Figure 1 shows a comparison of the two inference techniques (haplotype, genotype) as well as the two selection methodologies (greedy, random). Due to the non-negligible presence of some autozygosity within the cohort, these two distributions represent an upper and lower bound on the imputation capacity.

*Combined SNV calling*

We followed the protocol for best practice variant detection detailed in the GATK v2 documentation, with individual parameters tuned for low-pass data. The specific analysis steps are as follows:

1. Reads from all lanes for each sample were merged and duplicate molecules flagged.
2. For each sample, reads were locally realigned around small suspicious intervals (generally indels). For single nucleotide variants, dbSNP v130 and the SOLiD single-sample calling were used; for indels, dbSNP v130 and 1000 Genomes Project Pilot (07/2010) calls were used. Finally, mate pair reads were synchronized.
3. For each sample, base quality scores were recalibrated using per-base covariates: machine cycle for base; di-nucleotide combination for base; number of consecutive previous bases matching this base (accounting for homopolymers); position of the base in the read; the primer round for this base (SOLiD specific).
4. All samples were called together using the UnifiedGenotyper module, which uses an iterative Bayesian likelihood model to estimate allele frequency in the population and genotype calls. We allowed an emitted quality value of 10 and a minimum quality value of 30 for confident calls. Similarly, indels supported by at least 2 reads and at least 60% of all reads were also called for the purpose of masking. Any calls that were low confidence, overlapped a detected indel, or consisted of 3 or more SNVs within 10bp were excluded at this stage.
5. To assess novel variants as accurately as possible, we performed variant quality score re-calibration in three stages. We isolated variants at sites that were expected to be known using the HapMap calls (release 27), 1000 Genomes Project Pilot low-coverage calls (07/2010), and dbSNP (v129, downgraded to minimize new, poorer quality SNPs). We identified clusters within these calls based on quality, allelic balance, strand-bias, and homopolymer run. We then classified all variants according to their expected False Discovery Rate (FDR) given the established cluster boundaries and kept only those novel variants that had expected FDR below 20%.
6. Finally, we performed imputation of un-typed variants in two stages. First, we imputed variants within the cohort using the BEAGLE framework and incorporated any sites with a minimum $r^2$ cutoff of 0.50. Next, we imputed variants from the 1,000 Genomes (07/2010) haplotypes using MaCH. Processing each chromosome separately, we estimated model parameters from all seven samples and then used these estimates to perform a greedy imputation with default parameters, keeping any previously un-typed sites with a minimum $r^2$ cutoff of 0.50.

Our final set of calls consisted of high-specificity/low-sensitivity novel variants and aggressively called and imputed known variants. Variants not observed in dbSNP v130 were annotated as novel.

*Combined InDel calling*

Indels were called using the Dindel v1.01, a program for calling small indels from short-read sequence data. While Dindel does not yet explicitly model multiple independently sequenced samples, we performed the analysis in several rounds and shared the reference information across all samples. As per the user manual, we first generated a list of candidate indels and mate-pair distance distributions for each sample separately using the GATK realigned and recalibrated reads. We then pooled all of the candidate indels into a single reference library and mapped each individual against the pooled library to identify the final set of indels. We classified indels as previously known if they overlapped with any insertion/deletion site in dbSNP v130.

*Variant extrapolation*

We perform error estimates and variant extrapolation independently for each sample as well as separately for known and novel variants in accordance with the following protocol:

- For previously known variants we measure specificity and sensitivity based on the set of calls overlapping with the genotype array, taken as ground truth. Sensitivity is measured as the percentage of non-reference genotype sites that are called as non-reference in the sequence; specificity is measured as the percentage sequence sites called non-reference that are also called non-reference by the array.
- For novel variants, accurately measuring sensitivity is particularly difficult in low-pass data, and so we conservatively assume sensitivity to be the same as for known variants. For measuring specificity, we assume that the totality of calls is a mixture of true-positive variants with an expected transition bias and false-positive variants occurring randomly and exhibiting no transition bias[11]. Formally, given an expected transition rate of $\delta_{ex}$, an observed transition rate of $\delta_{ob}$, a true-positive percentage $\phi_{TP}$, and false-positive percentage $\phi_{FP}$ we establish the following system: $\phi_{TP} + \phi_{FP} = 1$ and $\delta_{ex}\phi_{TP} + \frac{1}{3}\phi_{TP} = \delta_{ob}$ and can solve for specificity as $\phi_{TP} = \frac{\frac{1}{3} - \delta_{ob}}{\frac{1}{3} - \delta_{ex}}$. We observe that this estimate is very consistent with the empirical specificity in known variants and novel variants based on experimental validation (Table 1).

- We take expected ratios of 2.10 for known sites and 2.07 for novel sites from the GATK variant detection best-practices (http://www.broadinstitute.org/gsa), calculated as weighted averages across the 1000 Genomes CEU and YRI trios.

For both variant types we then extrapolate the total expected number of variants in the standard way as expected = (observed) x (specificity) / (sensitivity).

**Figure Legends**

**Figure 1: Imputation strategy and information capacity in founder population**

**A:** Schematic outline of the strategy. IBD shared haplotypes (color-coded) are identified from genotype data (grey spots) in a cohort; a small panel of individuals with abundance of IBD is sequenced (top); and sequenced variants within shared haplotypes are inferred to the rest of the cohort within shared regions (bottom). **B:** Percentage of genomic content of the cohort that is inferred (Total Information Potential) as a function of sequenced reference panel size, calculated from autosomal genotype data. Rapid growth of Total Information Potential in the highly-related Kosraen population (2,906 samples, green) compared to slower growth in less related Ashkenazi Jewish population (1,200 Crohn's Disease case-control samples[27], brown).

**Figure 2: Population-specific genomewide allele frequency spectrum**

Variants previously observed in 1,000 Genomes Project pilot are plotted according to abundance in each sequenced genome (y-axis) as a function of allele frequency in the reference cohort (x-axis). Kosraen ("KOS") genome compared to sequenced Korean ("SJK"); Yoruban ("YRI"); and European ("JCV") genomes. Allele frequency spectrum measured in (**A**) East Asian - JPTCHB; (**B**) African - YRI; and (**C**) European - CEU origin reference cohorts.

**Figure 3: Frequency spectrum of putatively functional variants**

Histogram of variants annotated by their coding class and novelty. Each bar represents the percentage of variant sites (y-axis) at the respective allele frequency (x-axis) out of all observed in the respective variant class: (**A**) distribution of variants by functional class, showing significant enrichment for singleton nonsense mutations; (**B**) distribution of previously known (white) and novel (yellow) variants across the entire genome. Only sites where a call could be made in all samples are considered (total counts in Supplementary Table 1).

**Figure 4: Concordance of known and novel variants in IBD and non-IBD regions**

We examined concordance of called variants in previously predicted pairwise IBD regions. For all sites that are called homozygous in both samples with at least one being non-reference, we measure concordance (x-axis) as the percentage where both are non-reference. We expect 100% concordance in truly co-inherited regions with no sequence error. Y-axis shows percentage of IBD segments at a given concordance level. Concordance for previously known variants (**A**, white bar) and novel variants (**B**, yellow bar) is shown in comparison to randomly placed non-IBD regions of an equal length distribution (both, black points). On average, IBD segments maintained 99.6% (known) and 97.3% (novel) concordance compared 82.9% (known) and 31.0% (novel) in a background distribution of non-IBD segments.

**Supplementary Figure 1: Imputation capacity for genotype versus haplotype schemes**

Detailed breakdown of Total Information Potential as a function of sequencing budget for genotype-based (light blue) and haplotype-based (dark blue) inference; categorized according to random selection of samples (dashed line) and greedy optimization of sample selection (solid line).

**Supplementary Figure 2: Comparison of variants detected in published genomes**

Distribution of detected autosomal variants in previously published genomes shown compared to average Kosraen individual ("KOS (avg)" at left). Previously known variants (in dbSNP 130) shown in white bar and novel variants shown in yellow (extrapolation detailed in Materials & Methods).

**Supplementary Figure 3: Additional novel variants attained from further sequenced individuals**

Number of additional unique novel variant sites gained (y-axis) from each subsequent sequenced genome (x-axis). Counts were measured across variant sites observed in any sample (grey bar) and variant sites that could be called in all samples (yellow bar). The distribution represents an average across all unique permutations of the seven individuals.

**Supplementary Figure 4: Overall distribution of coding and general variants**

Each bar represents the percentage of variant sites (y-axis) at the respective allele frequency (x-axis) out of all sites observed in each variant class: all known (white), known coding (gray), all novel (yellow), novel coding (dark yellow). Only sites where a call could be made in all samples are considered (total counts in Supplementary Table 1).

**Supplementary Figure 5: Distribution of detected short insertions and deletions**

Histogram of novel (yellow) and previously known (white) indels across all seven individuals. Number of indels at each carrier level shown at cap of each bar.

**Supplementary Figure 6: Distribution of copy number variants detected**

Number and type of CNV identified in four sequenced individuals. Total number of discrete regions called (y-axis) as function of CNV length (x-axis).

**Supplementary Figure 7: Sequence concordance as function of IBS segment length**

Concordance of sequenced homozygous variants in array-based IBS regions of increasing length. Previously known (black line) and novel (yellow line) variants shown on left y-axis as a function of segment length on x-axis. Number of segments at each IBD length window shown in grey bars on right y-axis.

## Tables

**Table 1: Experimental validation of known and novel variants**

| Call type | Unique sites | Observed Ti/Tv ratio | Expected Ti/Tv ratio[1] | Calculated variant specificity[2] | Total calls validated | Experimental variant specificity[3] | Sensitivity |
|---|---|---|---|---|---|---|---|
| Known | 4,522,474 | 2.07 | 2.10 | 98.9% | 2,958,772 | 98.2%[4] | 92.8%[4] |
| Novel | 1,212,831 | 1.74 | 2.07 | 88.6% | 64 | 87.5% | - |

[1]From 1000 Genomes high-quality low-pass estimates.
[2]Calculated from difference in expected and observed Ti/Tv ratio (*Materials & Methods*).
[3]Percentage of non-reference calls experimentally validated as non-reference.
[4]Sensitivity/Specificity of calling SNPs of all seven samples together using GATK Unified Genotyper and imputation on combined low-coverage samples (Materials & Methods, *Combined SNV calling*).

**Supplementary Table 1: Read placement and coverage per sample**

**Supplementary Table 2: Array-based quality control in filtered sequenced variants**

**Supplementary Table 3: Experimental validation of called novel variants**

**Supplementary Table 4: Called and extrapolated variants in seven Kosraen samples**

**Supplementary Table 5: Putatively functional variants detected, called in all samples, and variant in a single sample**

# Figures

**Figure 1**

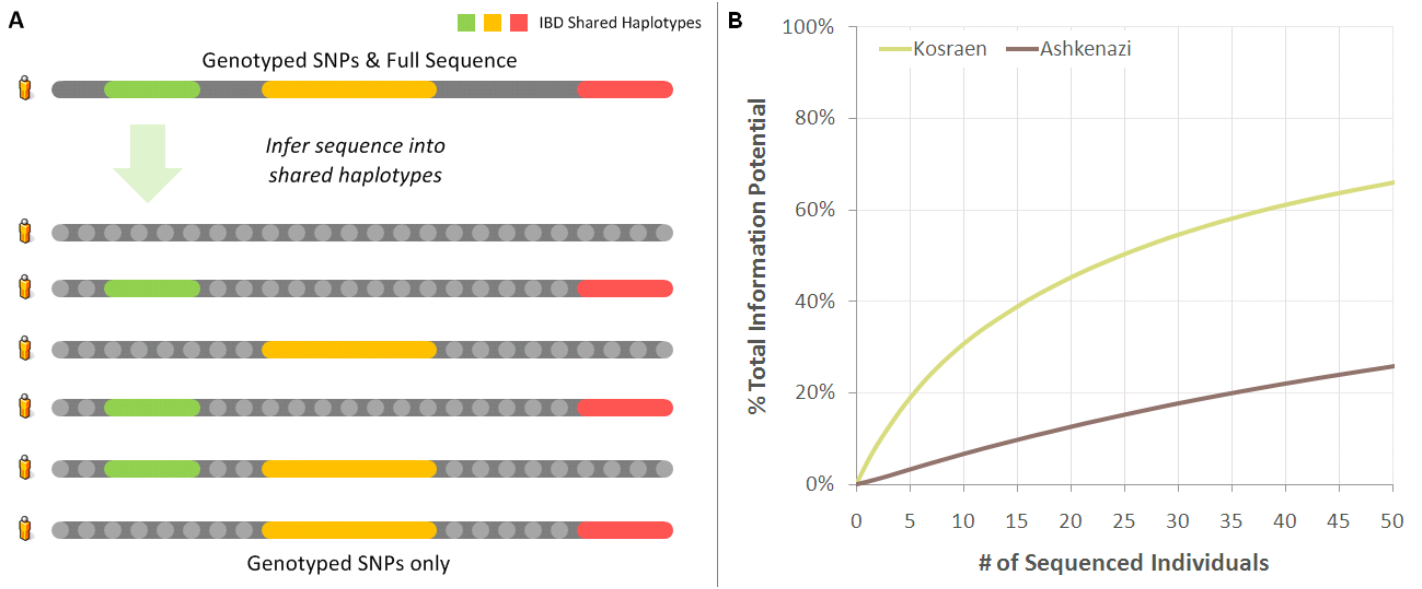

**Figure 2**

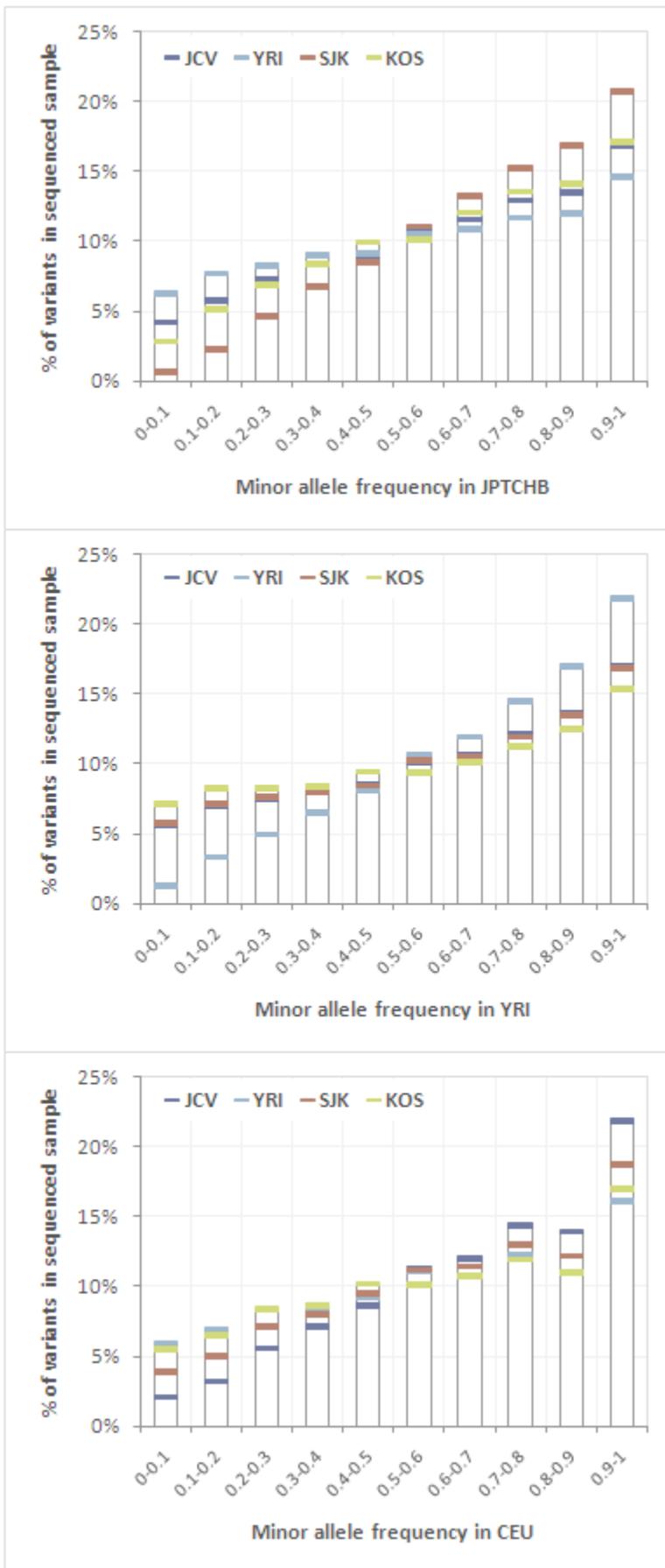

**Figure 3**

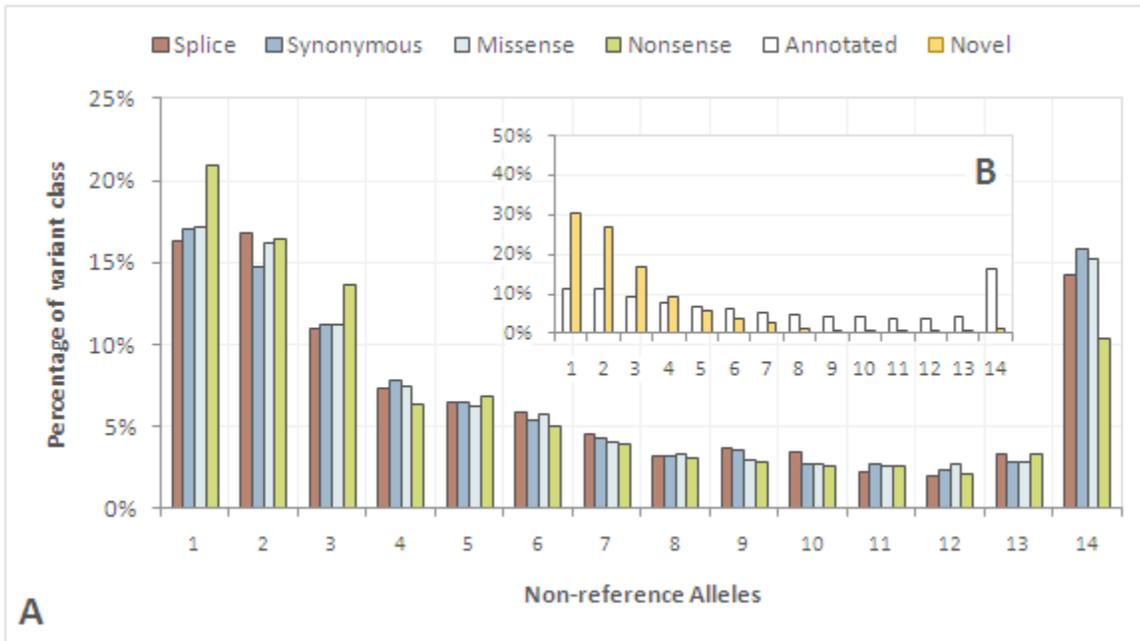

**Figure 4**

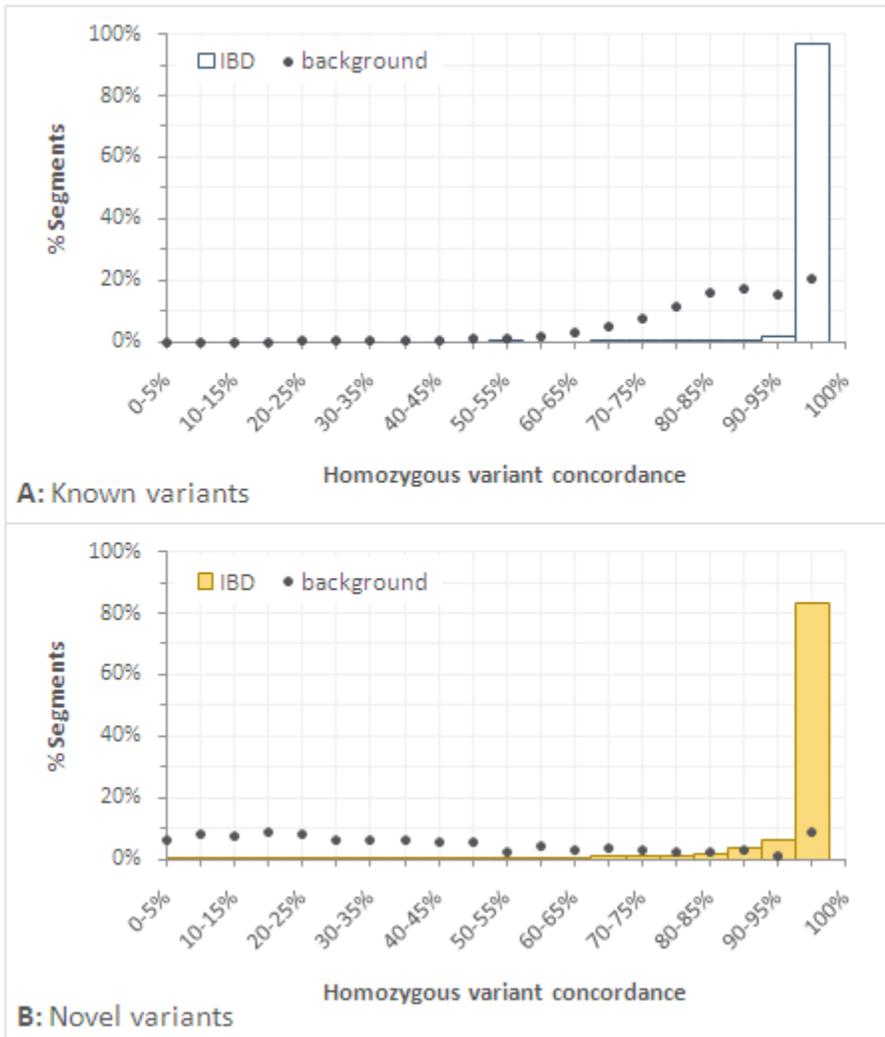

# Supplementary Figures and Tables

**Supplementary Figure 1**

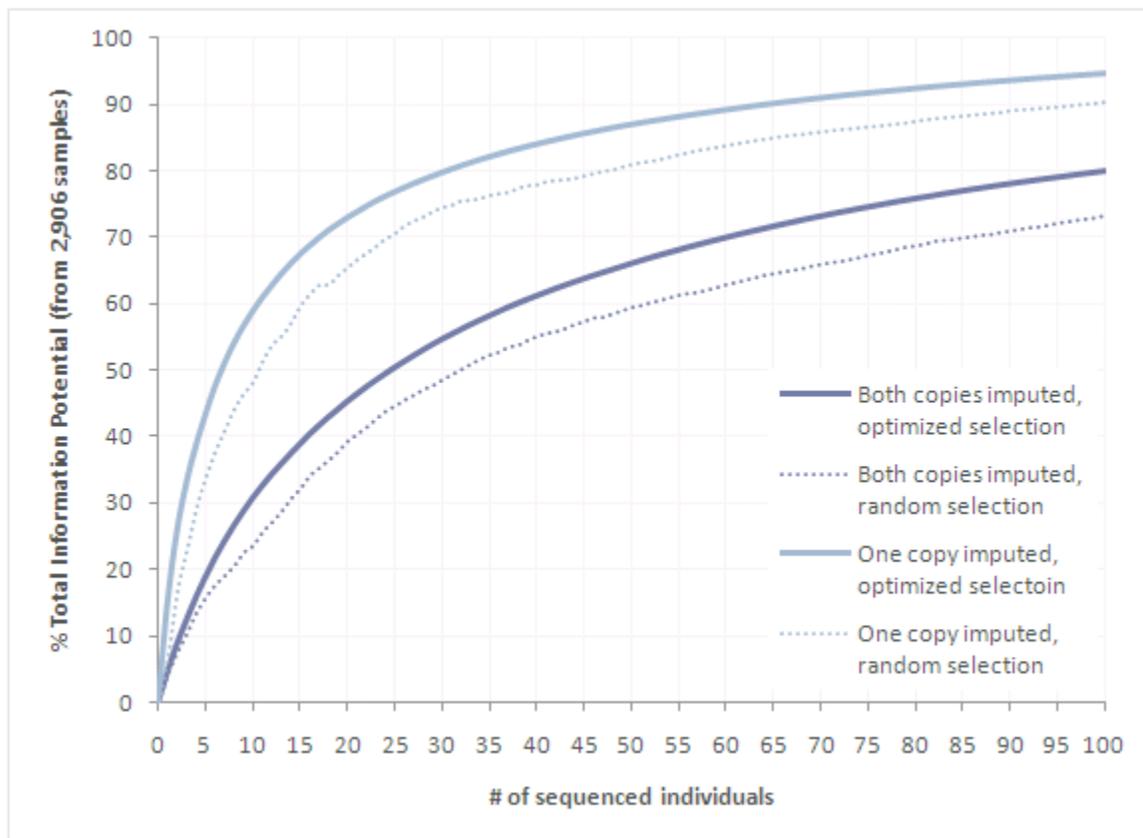

**Supplementary Figure 2**

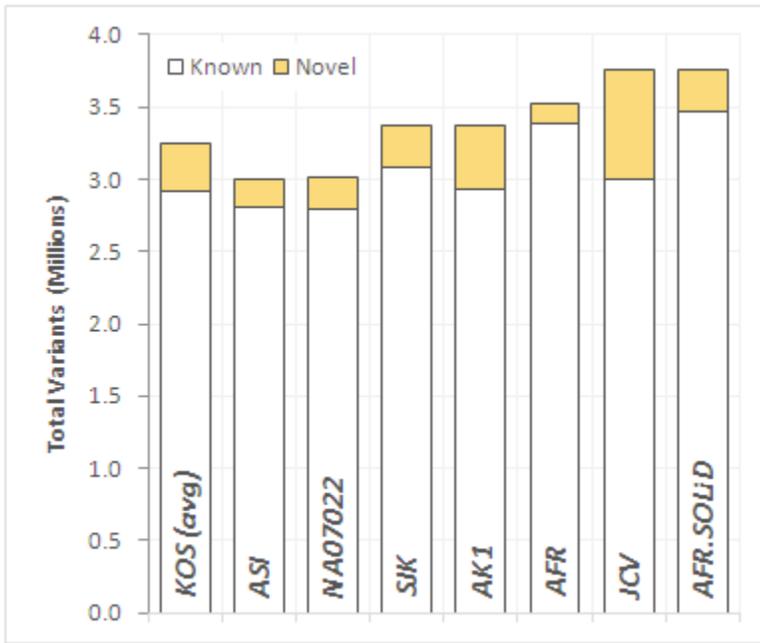

**Supplementary Figure 3**

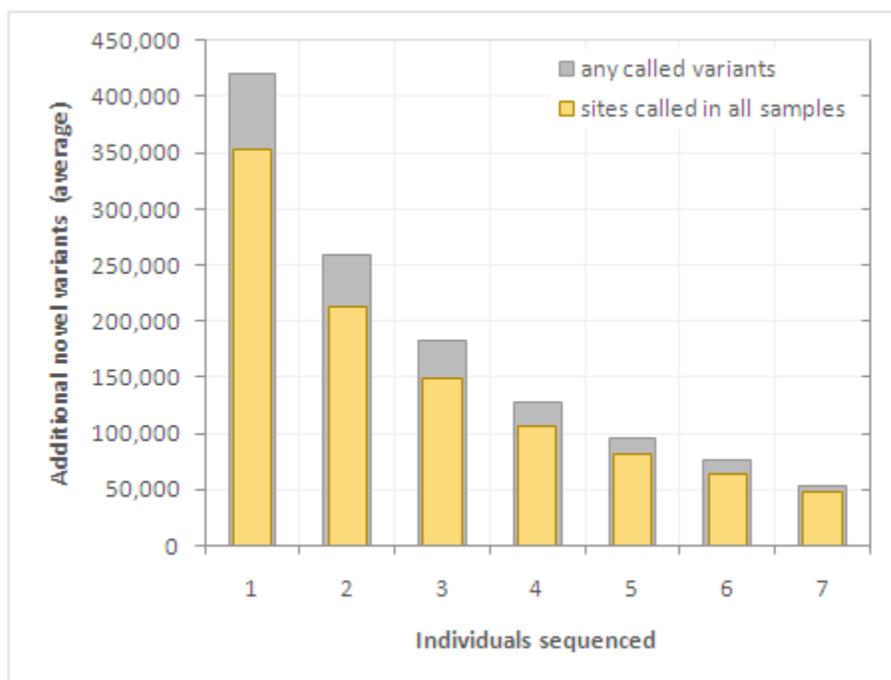

**Supplementary Figure 4**

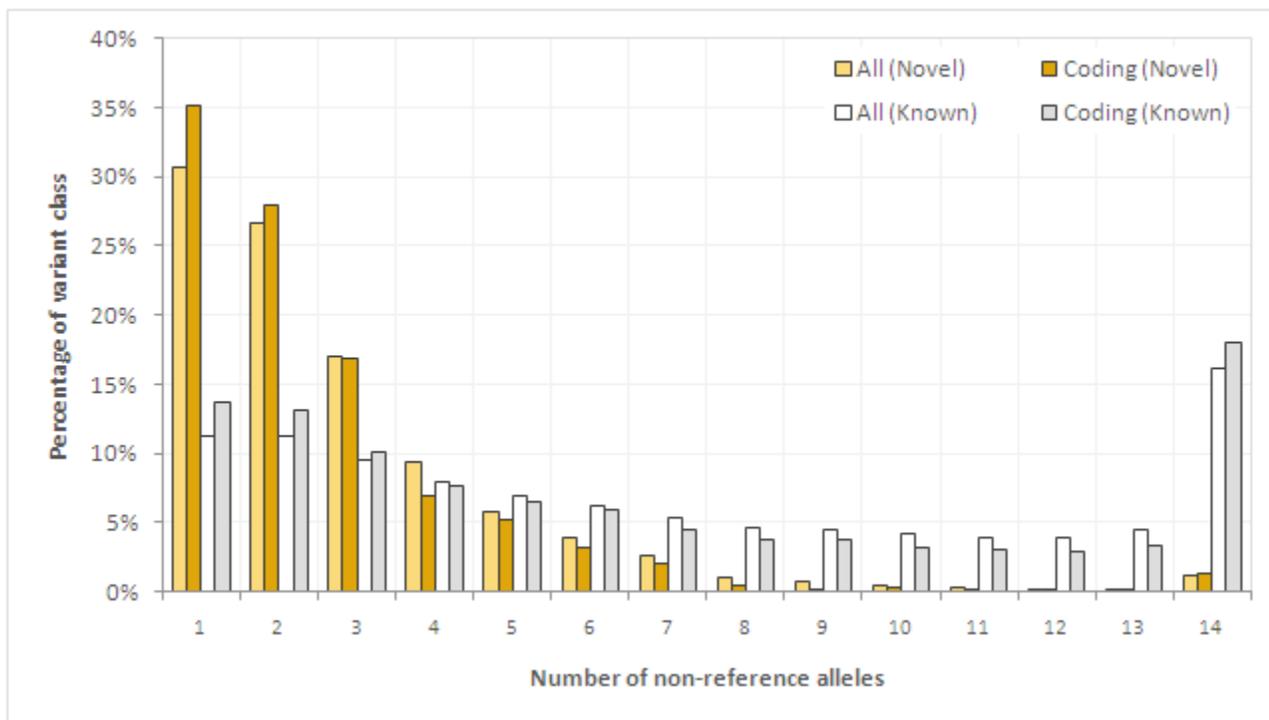

**Supplementary Figure 5**

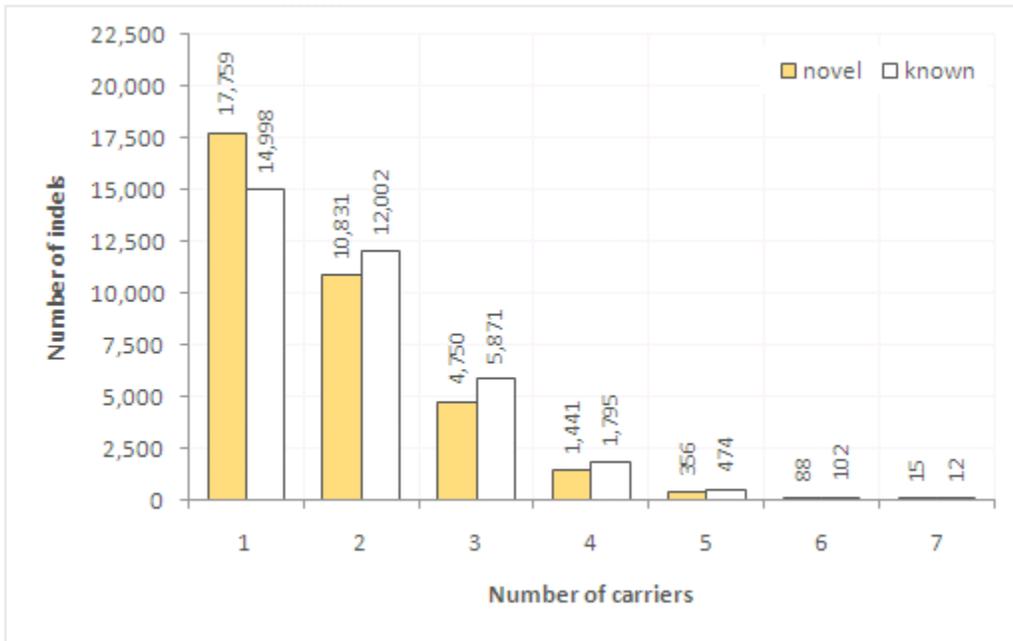

**Supplementary Figure 6**

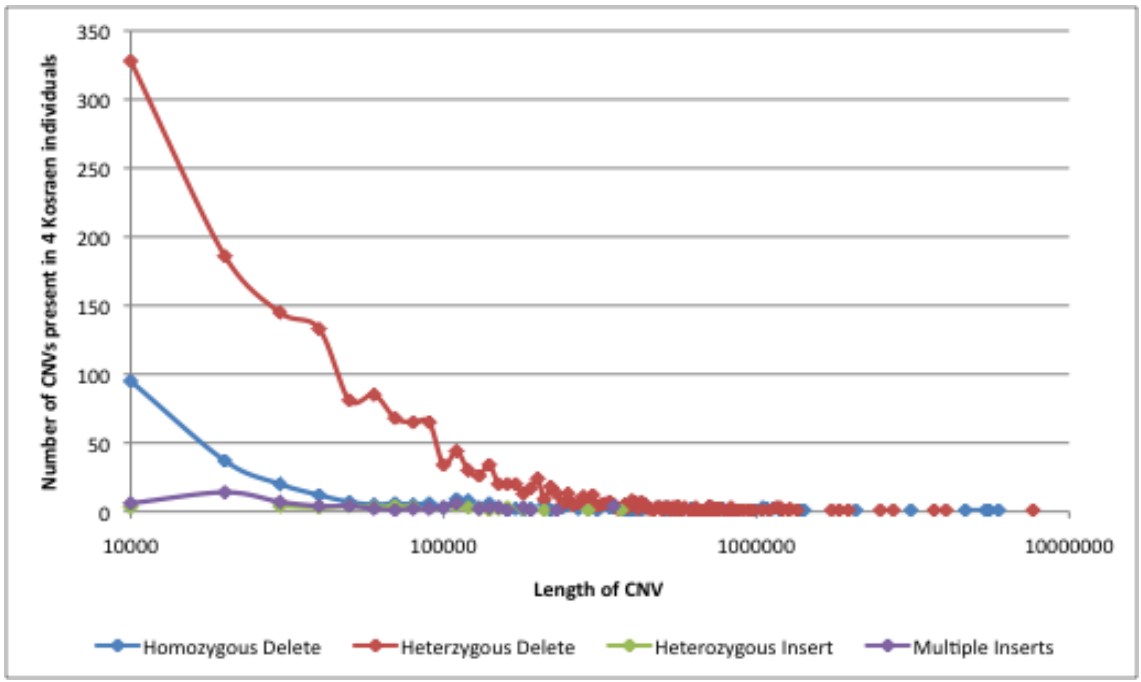

**Supplementary Figure 7**

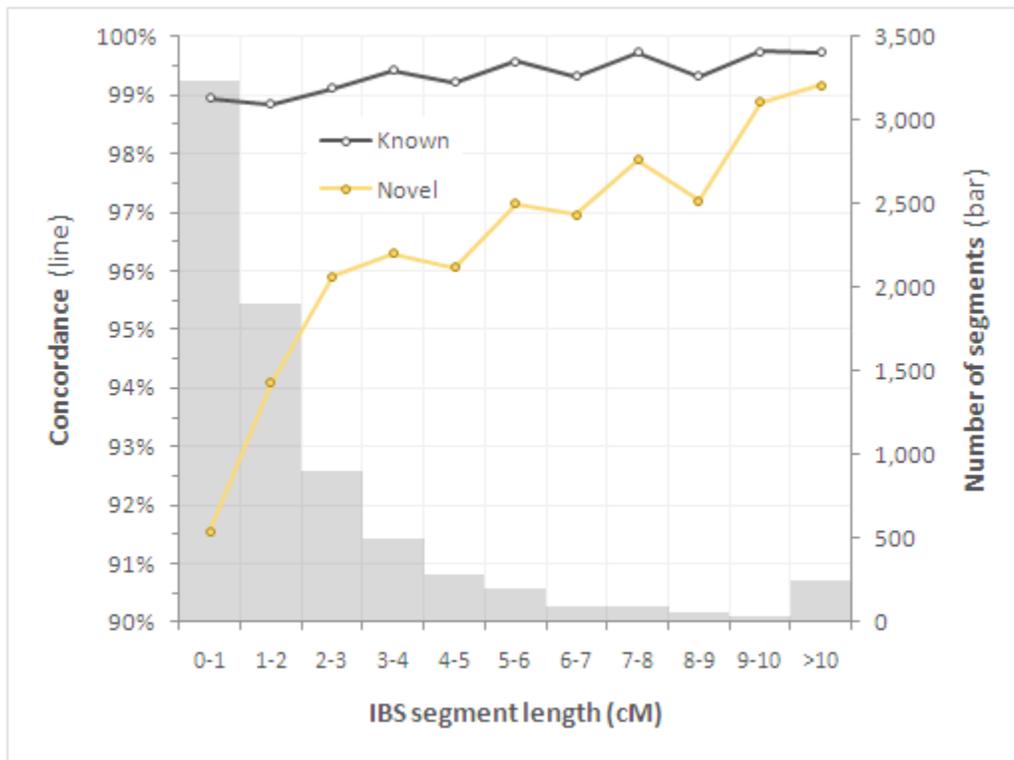

**Supplementary Table 1**

| ID | Read type | Raw reads | Aligned reads | Aligned mates | Non-redundant mates | Genomic Coverage |
|---|---|---|---|---|---|---|
| | 2x50 | 1,048,656,097 | 504,557,476 | 182,789,788 | 67,174,561 | |
| | 2x35 | 740,209,937 | 498,459,246 | 193,481,169 | 76,709,567 | |
| K1955 | sum | 1,788,866,034 | 1,003,016,722 | 376,270,957 | 143,884,128 | 4.32 |
| K2033 | 2x50 | 1,027,824,174 | 523,272,510 | 197,496,500 | 149,412,266 | 5.34 |
| K5866 | 2x50 | 1,012,198,246 | 558,930,954 | 220,603,222 | 123,969,777 | 4.43 |
| K1674 | 2x50 | 959,926,988 | 513,403,455 | 194,480,787 | 135,734,837 | 4.85 |
| K6169 | 2x50 | 1,786,661,548 | 671,728,313 | 225,801,169 | 169,854,108 | 6.07 |
| K6494 | 2x50 | 1,727,132,405 | 626,281,340 | 207,118,404 | 178,149,982 | 6.36 |
| K5675 | 2x50 | 676,989,864 | 298,728,860 | 215,139,732 | 87,855,892 | 3.14 |

Supplementary Table 2

| | | Single sample | | Raw multi-sample | | Filtered multi-sample | | BEAGLE internal | | MaCH: 1000 Genomes | |
|---|---|---|---|---|---|---|---|---|---|---|---|
| | Sample | Specificity | Sensitivity[1] | Specificity | Sensitivity[1] | Specificity | Sensitivity[1] | Specificity | Sensitivity[1] | Specificity | Sensitivity[1] |
| Reference | K6169 | 80.2% | 99.91% | 90.1% | 97.90% | 89.8% | 97.95% | 91.3% | 98.00% | 92.8% | 97.08% |
| | K2033 | 91.4% | 99.98% | 90.8% | 99.86% | 90.5% | 99.87% | 91.9% | 99.79% | 93.1% | 98.96% |
| | K5675 | 62.8% | 99.99% | 76.5% | 99.91% | 75.9% | 99.93% | 79.1% | 99.02% | 81.3% | 97.81% |
| | K1955 | 70.7% | 99.99% | 84.9% | 99.89% | 84.5% | 99.90% | 86.3% | 99.42% | 88.0% | 98.43% |
| | K1674 | 80.8% | 99.98% | 89.2% | 99.85% | 88.9% | 99.87% | 90.5% | 99.77% | 92.0% | 98.88% |
| | K5866 | 79.4% | 99.98% | 86.7% | 99.85% | 86.4% | 99.86% | 88.6% | 99.66% | 90.2% | 98.73% |
| | K6494 | 81.0% | 99.98% | 91.1% | 99.88% | 90.7% | 99.89% | 92.0% | 99.84% | 93.3% | 99.01% |
| | Sample | Specificity | Sensitivity | Specificity | Sensitivity | Specificity | Sensitivity | Specificity | Sensitivity | Specificity | Sensitivity |
| Heterozygous | K6169 | 99.1% | 49.8% | 88.5% | 82.3% | 88.5% | 81.1% | 89.2% | 83.8% | 87.9% | 87.4% |
| | K2033 | 99.9% | 72.5% | 98.2% | 79.1% | 98.2% | 77.6% | 98.2% | 81.1% | 96.3% | 85.1% |
| | K5675 | 99.8% | 7.4% | 98.3% | 36.1% | 98.5% | 35.0% | 94.8% | 53.0% | 91.5% | 57.5% |
| | K1955 | 99.9% | 21.4% | 98.4% | 60.1% | 98.4% | 58.5% | 97.0% | 68.6% | 94.7% | 73.3% |
| | K1674 | 99.9% | 44.8% | 98.1% | 77.4% | 98.1% | 75.7% | 98.0% | 79.8% | 96.1% | 84.0% |
| | K5866 | 99.9% | 38.2% | 98.4% | 70.3% | 98.4% | 68.5% | 97.8% | 74.7% | 95.8% | 79.5% |
| | K6494 | 99.9% | 54.4% | 97.8% | 82.4% | 97.8% | 81.3% | 98.0% | 84.2% | 96.2% | 87.4% |
| | Sample | Specificity | Sensitivity | Specificity | Sensitivity | Specificity | Sensitivity | Specificity | Sensitivity | Specificity | Sensitivity |
| Homozygous Variant | K6169 | 88.0% | 83.2% | 95.0% | 91.5% | 95.0% | 91.3% | 95.9% | 93.1% | 95.6% | 93.2% |
| | K2033 | 92.8% | 96.8% | 93.9% | 95.8% | 93.9% | 95.6% | 95.0% | 97.5% | 94.7% | 97.7% |
| | K5675 | 77.8% | 47.7% | 80.9% | 80.1% | 81.0% | 79.9% | 85.2% | 94.0% | 84.7% | 94.4% |
| | K1955 | 80.6% | 68.0% | 87.9% | 88.9% | 88.0% | 88.6% | 90.2% | 96.2% | 89.8% | 96.4% |
| | K1674 | 85.6% | 86.5% | 93.2% | 95.6% | 93.2% | 95.3% | 94.6% | 97.3% | 94.3% | 97.5% |
| | K5866 | 83.2% | 86.2% | 90.4% | 94.6% | 90.5% | 94.3% | 92.5% | 97.1% | 92.2% | 97.3% |
| | K6494 | 90.1% | 82.2% | 95.4% | 95.5% | 95.4% | 95.3% | 96.3% | 97.2% | 96.0% | 97.3% |

1: Sensitivity of calling SNPs of all seven samples together measured as percentage of genotyped reference alleles called as non-reference or no-call

**Supplementary Table 3**

|  | Total | Validated Ref | Validated Het | Validated Var | Call specificity | Non reference specificity |
|---|---|---|---|---|---|---|
| Called Het. | 28 | 4 | 24 | 0 | 86% | 87.5% |
| Called Hom. | 36 | 4 | 14 | 18 | 50% | |

# Supplementary Table 4

| Sample | Total Called | Known hom | Known het | Novel hom | Novel het | Total expected[1] |
|---|---|---|---|---|---|---|
| K1674 | 3,241,805 | 1,587,554 | 1,227,402 | 88,395 | 338,454 | 3,282,447 |
| K1955 | 3,082,651 | 1,628,324 | 1,073,233 | 87,563 | 293,531 | 3,214,836 |
| K2033 | 3,213,192 | 1,606,658 | 1,187,234 | 90,701 | 328,599 | 3,232,382 |
| K5675 | 2,814,993 | 1,638,929 | 841,377 | 91,277 | 243,410 | 3,106,023 |
| K5866 | 3,197,269 | 1,612,003 | 1,177,740 | 91,720 | 315,806 | 3,266,299 |
| K6169 | 3,389,746 | 1,500,652 | 1,393,928 | 80,226 | 414,940 | 3,304,190 |
| K6494 | 3,281,503 | 1,566,269 | 1,243,493 | 87,240 | 384,501 | 3,281,031 |

[1]Extrapolated from known/novel variant error rates (Supplemetary Methods)

**Supplementary Table 5**

| Variant type: | Unique sites | Called in all | Singletons |
|---|---|---|---|
| All | 5,735,305 | 5,035,424 | 15.2% |
| Annotated | 4,522,474 | 4,021,114 | 11.3% |
| Novel | 1,212,831 | 1,014,310 | 30.6% |
| All coding | 31,229 | 26,281 | 17.2% |
| Synonymous | 14,659 | 12,219 | 17.0% |
| Missense | 15,843 | 13,447 | 17.1% |
| Nonsense | 751 | 615 | 21.0% |
| Splice variants | 2,785 | 2,356 | 16.3% |

# Supplementary Methods

## Preliminaries

### Terminology and Notation

*Identity by Descent (IBD).* A pair of descendants of the same ancestor is IBD if they share haplotypes that have been transmitted along the respective lineages leading to them. The shared haplotypes lie on homologous chromosomes of different individuals or of the same individual, in the latter case the individual has related parents. Let $P = \{1, 2, …, n\}$ be the set of all individuals belonging to the population under study. We denote by $R(i,j)$ the collection of shared regions between individuals $i, j \in P$. A shared region in $R(i,j)$ is identified by a triplet $(l,r,c)$, where $l$ is its left endpoint, $r$ right endpoint and $c \in C = \{1, 2, …, 22\}$ is the relevant chromosome.

*Total information Content.* Our aim is to sequence only a subset of individuals to infer information about the unsequenced population. *Total Information Content* (*TIC*) of a set $Q$ is the fraction of the cohort members' genomes that we directly obtain or indirectly can infer by sequencing a subset of individuals. Formally, if each unsequenced individual $i$ has regions of total length $L_i$ of her $L$-long genome inferred and $L_i(Q)$ as , then

$$TIC(Q) = \frac{|Q|L + \sum_i L_i(Q)}{|P|L}.$$

*Utility of Sequencing an Individual.* Given a set of already sequenced individuals $Q$, we associate each individual $i \in P\backslash Q$ with it a quantity $U(i,Q)$ that corresponds to the utility of sequencing $i$ at this stage. $U(i,Q)$ is the total length of regions that $i$ shares with all unsequenced individuals across all chromosomes i.e.

.

*Interval trees.* Our method relies on an interval tree data structure. An interval tree is an ordered and self-balancing tree data structure. We have an interval tree $t(i,c)$ for each individual $i$ and each chromosome $c$.

## Problem Definition

We define the problem of Representative Selection as follows:

**Input:**  The function $R$ listing sets of shared regions for each pair of samples in population $P$

  Total budget $b$

**Output:** Set $Q \subset P$ of $b$ individuals to sequence such that TIC($Q$) is maximized

Our proposed methodology helps to determine which subset of individuals to sequence. The size of the subset is the sequencing budget $b$ and the objective is to maximize *TIC*. This problem is reducible to the classic NP-hard optimization problem of Vertex Cover (VC): A special case where individuals either share the entire genome or nothing. Each vertex represents an individual and shared region of an individual on a particular chromosome. The edges represent the relatedness between regions. The problem then becomes that of picking a set of shared regions (individuals) such that we cover maximum number of edges.

**Algorithm**

**Informal Outline**

We propose a greedy approach, selecting individuals one at a time, gradually admitting samples into the set $Q$. **Error! Reference source not found.** shows the formulation for this approach.

1) While $|Q| \leq b$
   a) Find the individual $j \in I$ such that $U(j,Q) = \max_{i \in P\backslash Q} U(i,Q)$. $Q \leftarrow Q \cup \{j\}$
   b) Exclude complete regions $(l,r,c)$ in $R(i,j)$, $i \in P$ and all parts of regions $(l',r',c)$ in $R(i,k)$, $k \in P\backslash\{j\}$ that overlap with $(l,r,c)$ in $R(i,j)$

**Algorithm 1: Greedy Method for picking budget b number of individuals**

**Data structure details**

Naive implementation of this greedy approach runs into the computational bottleneck of maintaining lists of shared regions for each pair of individuals. Intuitively, such regions keep getting shattered by interval exclusion operation in step 2 of **Error! Reference source not found.**. Efficient implementation that maintains these intervals requires a special data structure. We use interval trees for this purpose. Each node in the tree contains an interval representing a shared region along with a pointer to the node in the tree of the other individual with whom the region is shared. The first step is to calculate $U(i,Q)$, for each individual $i \in P\backslash Q$. Our greedy approach now selects the individual $j$ with the highest value of $U(i,Q)$. Before we make the next choice, we need to exclude regions that have been imputed by picking $j$. These are complete regions $(l,r,c)$ in $R(i,j)$, $i \in P$ which we will impute directly by sequencing individual $j$. Additionally, we will indirectly impute parts of regions $(l',r',c')$ in $R(i,k)$, $k \in P\backslash j$ that overlap with $(l,r,c)$ in $R(i,j)$. We then recalculate $U(i,Q)$ $i \in P\backslash j$ and make the next greedy choice followed by elimination of newly imputed segments. This continues till we have picked individuals up to our sequencing budget $b$. To understand this, consider a simple example. Suppose individuals A and B share region (5,20,1); B and C share (13,25,1) and A and C share (30,50,1). We see that $U(A,Q) = 35$, $U(B,Q) = 27$, $U(C,Q) = 32$. The greedy algorithm will first pick A and add it to Q. We exclude directly imputed regions (5,20,1) of R(A,B) and (30,50,1) of R(A,C). Also, the region (13,20,1) in R(B,C) has been indirectly imputed by picking A. The recalculated $U(B,Q) = U(C,Q) = 5$.

**Comparison to linked lists**

The individuals in one of the data sets we considered shared on an average thirteen million regions per chromosome. The magnitude of the dataset demands efficient management of these shared regions. Operations on regions like insertion into the interval tree, deletion, querying for overlaps (for finding intersection with imputed regions) and modifications need to be done quickly. Thus, a data structure was required that optimally scales to these large number of regions. One of the possible choices was a linked list. The linked list provides speed in terms of construction ($O(n)$ time), insertion and deletion ($O(1)$ time) of segments with linear and constant running times respectively. But the bottleneck was querying for overlaps ($O(n)$ time for linked list) and subsequent modification of the regions, where majority of the running time was spent. In some cases the entire linked list may have to be traversed before an overlap is found and the worst case running time is linear in the number of shared regions. Thus, the linked list is not the best solution for this implementation. The interval tree data structure was a better alternative. Since a region is represented with a start and end point, it was suitable to be modeled as an interval or node of the tree. Moreover the worst case time to query for an overlap (see Fig. 3.) required $O(\log n + m)$ time, where the $n$ refers to the total number of intervals in the tree and $m$ refers to the number of overlapping intervals in the query. This improvement in running time is due to the balanced nature of the interval tree that made it possible to search in only a section of the tree to find the overlapping intervals.

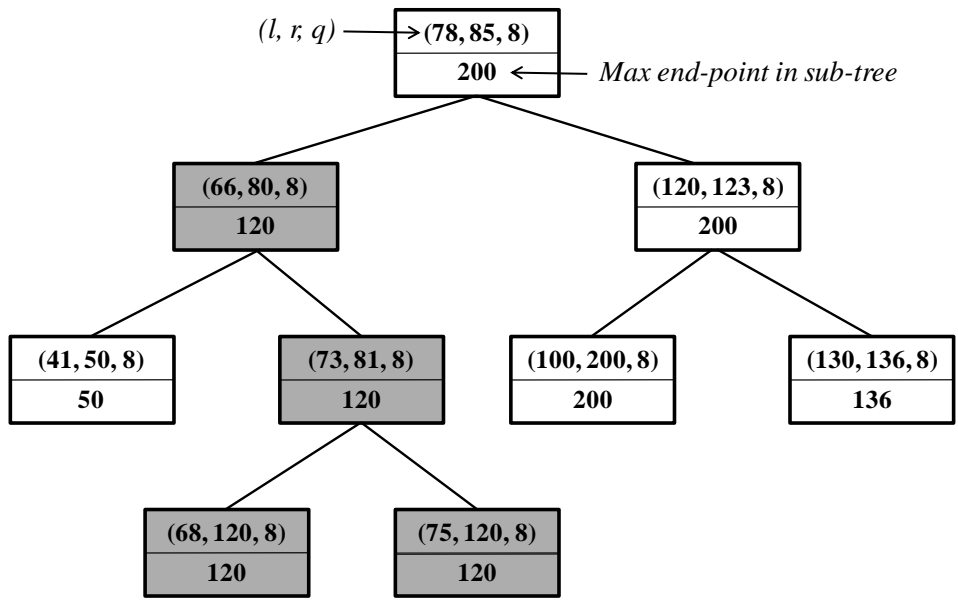

*Fig. 3. Representation of Interval Tree. Each node represents a shared region. The highlighted intervals are returned if we were to query for intervals overlapping (66,78,8).*

**Formal Algorithm**

```
ALGORITHM TREE-CREATION (T, R):

1) For all c ∈ C
   a) For all (l,r,c) in R(i,j) ∈ R
      i)  Insert node (l,r,c) into t(i,c) and t(j,c)
      ii) Set pointers between intervals (l,r,c) in t(i,c) and t(j,c)
```

**Algorithm 2: TREE-CREATION creates an interval tree for each individual i and each chromosome c and sets pointers between intervals shared by pairs of individuals.**

```
ALGORITHM PICK-INDIVIDUAL (T):

1) For all i ∈ P\Q
   a) Calculate U(i,Q)
2) Pick j | U(j,Q) = max_{i ∈ P\Q} U(i,Q). Q ← Q ∪ { j }
3) Delete all (l,r,c) in R(i,j) from t(j,c) and t(i,c) ∀ c ∈ C, i ∈ P\Q
```

**Algorithm 3: PICK-INDIVIDUAL picks individual j with the highest utility value and deletes intervals directly imputed by picking j.**

```
ALGORITHM INTERVAL-MODIFICATION (j,T):

1) For all c ∈ C
   a) Create list Ls(j,c) = { i ∈ P\Q | ∃ (l,r,c) in R(i,j) }
   b) For all i ∈ Ls(j,c)
      i) For all (l,r,c) in R(i,j)
         (1) For each (l',r',c) ∈ t(i,c) such that the intervals (l,r) and (l',r') overlap. Let k
             be such that (l',r',c) ∈ R(i,k)
             (a) Case 1:
                 (i) If l'≥l and r'≤r
                     1. Delete (l',r',c) from t(i,c) and t(k,c)
             (b) Case 2:
                 (i) If (l'<l and r'≤r)
                     1. Delete (l',r',c) from t(i,c) and t(k,c)
                     2. Insert (l',l,c) in t(i,c) and t(k,c)
                 (ii) Else if (l'≥l and r'>r)
                     1. Delete (l',r',c) from t(i,c) and t(k,c)
                     2. Insert (r,r',c) in t(i,c) and t(k,c)
             (c) Case 3:
                 (i) If (l'< l and r' > r)
                     1. Delete (l',r',c) from t(i,c) and t(k,c)
                     2. Insert (l',l,c) and (r,r',c) in t(i,c) and t(k,c)
```

**Algorithm 4:** SEGMENT-MODIFICATION eliminates parts of segments that are indirectly imputed by picking individual j. It first creates a list of individuals sharing segments with j, traverses their interval trees checking for three possible cases of overlaps and makes suitable modifications.

**Complexity**

To determine the complexity of the algorithm we need to take into account the two main operations; construction of the interval tree and querying for overlap. Construction of an interval tree requires $O(n \log n)$ time, where $n$ represents the average number of shared regions per individual. Querying for overlap requires $O(\log n + m)$ time, with $n$ being the total number of intervals in the interval tree and $m$ being the number of overlapping intervals. Thus, the total complexity can be given as $O(n \log n + \log n + m)$. The running time scales linearly in the number of shared regions. In terms of space complexity the interval tree requires $O(n)$ space.

**Implementation**

The algorithm was implemented in C++ and is made available for download at http://www1.cs.columbia.edu/~itsik/INFOSTIP/readme.html. All experiments were conducted on Linux-based cluster controlled by Sun Grid Engine on a node with 16 GB memory.